\documentstyle[prl,twocolumn,aps,epsf]{revtex}

\begin{document}

\title{ \bf Polaron Problem by Diagrammatic Quantum Monte Carlo }

\author{
Nikolai V. Prokof'ev and Boris V. Svistunov}

\address{Russian Research Center
``Kurchatov Institute", 123182 Moscow, Russia }

\maketitle

\begin{abstract}
We present a precise solution of the polaron problem by a novel
Monte Carlo method. Basing on conventional diagrammatic expansion
for the Green function of the polaron, $G({\bf k}, \tau)$, 
we construct a process of generating continuous random variables 
${\bf k}$ and $\tau$, with the distribution function exactly 
coinciding with $G({\bf k}, \tau)$. The polaron spectrum is
extracted from the asymptotic behavior of the Green function.
We compare our results for the polaron energy 
with the variational treatment of Feynman, and for the first time 
present precise dispersion curve which features an ending point 
at finite momentum.
\end{abstract}

\bigskip
\noindent PACS numbers: 71.38.+i, 02.70.Lq, 05.20.-y 
\bigskip

The polaron problem has a very long history starting from the work
by Landau \cite{Landau33}. In the most general form it is formulated as
what happens to the particle when it is coupled to the environment, and 
what are the properties of the resulting object, called polaron,
which consists of the bare particle dressed by environmental excitations.
This problem arises over and over again not only because
it is of fundamental importance both for the high-energy and for the 
condensed matter physics, but also because the notion of what we call
``particles'' becomes more diverse and new kinds of environments
appear. In this paper
we describe how the polaron problem can be solved numerically 
without systematic errrrs
using diagrammatic Monte Carlo method, and present the solution of
the notorious Fr\"{o}hlich model (see, e.g., 
Refs.\ \cite{Frohlich,Feynman-book}).
First, we explain in detail how this model fits into the general 
Monte Carlo scheme \cite{our96,our98} dealing with distribution functions
of continuous variables. We then describe the procedure of extracting the
polaron spectrum, $E(k)$, from the asymptotic decay of the Green function.
Although for small electron-phonon couplings the
polaron energy, $E_0$, and the effective mass, $m_*$,
i.e., the bottom of the polaron band,  are 
rather well given by the perturbation theory, the  perturbative 
approach fails to describe the spectrum near the threshold $E(k)-E_0
\approx \omega_p$, where $\omega_p$ is the frequency of the optical phonon.
In fact, the threshold features an ending point \cite{Whitfield,Larsen}
\begin{equation}
E(k)=E_0+\omega_p-{(k-k_c)^2 \over 2m_c} ~~~~~~~~ (k<k_c)\;,
\label{ending-point}
\end{equation}
analogous to the ending point of the excitation spectrum in $^4$He
described by Pitaevskii \cite{Pitaevskii}. Our numerical data
unambiguously confirm this conclusion. 

We start by considering the underlying mathematics. 
Suppose that for a certain
random variable/set of variables, $y$, the distribution function 
,$Q(y)$, is given in terms of a series of integrals with ever increasing
number of integration variables:
\begin{equation}
Q(y) \, = \, \sum_{m=0}^{\infty} \sum_{\xi_m} \int
dx_1 \cdots dx_m \, F(\xi_m, y, x_1, \ldots , x_m) \; .
\label{main}
\end{equation}

Here $\xi_m$ indexes different terms of the same order $m$. The
term $m=0$ is understood as a certain function of $y$. 
In Refs.\ \cite{our96,our98}
it was shown how to arrange a Metropolis-type stochastic process
simulating the distribution $Q(y)$ {\it exactly}. The process has very 
much in common with the Monte Carlo simulation of a distribution
given by a multi-dimensional integral. Nevertheless, there is an
essential difference associated with the fact that integration
multiplicity in the expansion Eq.\ (\ref{main}) is varying. 

\begin{figure}
\begin{center}
\epsfxsize=0.15\textwidth
\epsfbox{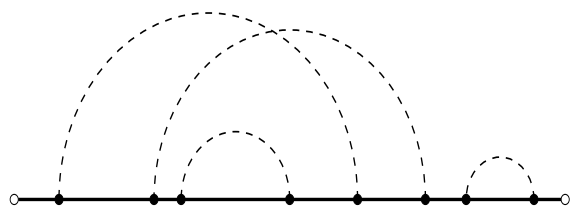}
\end{center}
\caption{A typical  diagram contributing to 
the polaron Green function.}
\label{fig:diagram}
\end{figure}

Projection onto the polaron problem is as follows. Let us interpret
the Matsubara (imaginary time) Green function of the polaron in 
the momentum-time representation, $G({\bf k}, \tau)$, as the distribution
function for the random variables ${\bf k}$ and $\tau$.  
We thus identify $G$ with $Q$, and $({\bf k}, \tau )$ with $y$. 
Equation (\ref{main}) is then identified with the
diagrammatic expansion of $G({\bf k}, \tau)$ in terms of free-electron 
and phonon propagators within the framework of conventional Matsubara
technique at $T=0$. Then, the variables $x_1, x_2, \ldots , x_m$
are the internal times and independent momenta of the diagram 
$\xi_m$. A typical diagram is presented
in Fig.\ 1. Solid lines denote the free-electron propagators, 
$G^{(0)}({\bf p}, \tau_2 - \tau_1 ) = 
\exp [-(p^2/2-\mu )(\tau_2 - \tau_1)]$, where $\mu$ is
the chemical potential. (Plank's constant and electron 
mass are set equal to unity).
Dashed lines and points stand for phonon propagators,
$D({\bf q}, \tau_2 - \tau_1 )$, and vertexes of the
electron-phonon coupling, $V({\bf q})$, respectively. Without
loss of generality, one may fix the left end of the diagram
at the origin of imaginary time, ascribing thus the time
$\tau$ to the right end.

In this paper we confine ourselves to the Fr\"{o}hlich model \cite{Frohlich}
where phonons are considered to be dispersionless, and the 
electron-phonon coupling has the form 
\begin{equation}
H_{\mbox{\scriptsize e-ph}} \, = \, 
\sum_{{\bf k},{\bf q}} V({\bf q}) \, 
\left( b^{\dag}_{\bf q} - b_{-{\bf q}}   \right) \,
a^{\dag}_{{\bf k}-{\bf q}} a_{\bf k} \; ,
\label{e-ph}
\end{equation}
\begin{equation}
V({\bf q}) \, = \, i \, \left( 2 \sqrt{2} \alpha \pi \right)^{1/2} \,  
\frac{1}{q} \; .
\label{V}
\end{equation}
In Eq.\ (\ref{e-ph}) $a_{\bf k}$ and $b_{\bf q}$ are the annihilation 
operators for the electron with momentum $\bf k$ and for the phonon 
with momentum $\bf q$, respectively; $\alpha$ is a dimensionless
coupling constant. In the Fr\"{o}hlich model
the phonon propagator is independent of momentum: 
$D({\bf q}, \tau_2 - \tau_1 ) = 
\exp [- \omega_p (\tau_2 - \tau_1)]$. It is convenient, however, to attribute the vertex
factors to the dashed lines, so that a dashed line with the momentum
$\bf q$ contributes the factor 
$\tilde{D}({\bf q}, \tau_2 - \tau_1 ) =  
\left| V({\bf q}) \right|^2 D(\tau_2 - \tau_1)$ to the diagram. 
The function $F$ is thus expressed as a
product of $G^{(0)}$'s and $\tilde{D}$'s, in accordance with
the standard diagrammatic rules.

Simulating the distribution $Q(y)$ is the process
of sequential stochastic generation of diagrams, identical to
functions $F$. (In our case these are the diagrams like that 
in Fig.\ 1, with certain fixed times and momenta.) The global 
process is constituted by a number of elementary sub-processes 
falling into two qualitatively different classes: (I) those which 
do not change the type of the diagram (change the values of variables
of corresponding function $F$, but not the function itself),
and (II) those which do change the structure of the diagram.
The processes of the class I are rather straightforward, being 
identical to those of simulating continuous distribution 
corresponding to the given function $F$. In this paper we use only
one process of this type, namely, shifting in time the right end of
the diagram Fig.\ 1.

In the heart of the method are the sub-processes of type II.
The generic rules for constructing them are as follows \cite{our98}.
Suppose a certain sub-process $\cal A$ transforms a diagram
$F(\xi_m, y, x_1, \ldots , x_m)$ into 
$F(\xi_{m+n}, y, x_1,  \ldots , x_m, x_{m+1}, \ldots , x_{m+n})$,
and, correspondingly, its counterpart $\cal B$ performs the inverse
transformation. For $n$ new variables we introduce vector
notation: ${\vec x} = \{ x_{m+1}, x_{m+2}, \ldots ,  x_{m+n} \}$.
The process $\cal A$ involves two steps. First, it {\it proposes} 
a change, selecting a new type of diagram, $\xi_{m+n}$, and a
particular value of $\vec x$. The vector $\vec x$ is selected
with a certain distribution function $W(\vec x)$. There are no 
requirements strictly fixing the form of $W(\vec x)$, 
but to render the algorithm most efficient, it is desirable that 
$W(\vec x)$ be chosen in accordance with some {\it a priori} 
knowledge of coarse-grained statistics of the vector $\vec x$. 
Upon proposing the modification, the process $\cal A$ either 
accepts it, with probability, $P_{\mbox{\scriptsize acc}}(\vec x)$, 
or rejects. The process $\cal B$ either accepts a modification, 
removing variables $\vec x$ (with a certain probability 
$P_{\mbox{\scriptsize rem}}(\vec x)$), or rejects this
modification.
For the pair of complimentary sup-processes to be balanced, the 
following Metropolis-like prescription should be fulfilled 
\cite{our98}:
\begin{equation}
P_{\mbox{\scriptsize acc}}(\vec x)  \: = \: \left\{ 
\begin{array}{ll}
R(\vec x) / W(\vec x)  ,
\mbox{~~~~~~~ if $R(\vec x) <  W(\vec x) $} \; , \\
\;\;\;\;\;\;1 \; ,  \mbox{~~~~~~~~~~~~~~~~ otherwise} \;\;\; ,
\end{array} \right.
\label{P_acc}
\end{equation}
\begin{equation}
P_{\mbox{\scriptsize rem}}(\vec x)  \: = \: \left\{ 
\begin{array}{ll}
W(\vec x) / R(\vec x)  ,
\mbox{~~~~~~~ if $R(\vec x) > W(\vec x) $} \; , \\
\;\;\;\;\;\;1 \; ,  \mbox{~~~~~~~~~~~~~~~~ otherwise} \;\;\; ,
\end{array} \right.
\label{P_rem}
\end{equation}
where
\begin{equation}
R(\vec x) \, = \, \frac{p_{\cal B}}{p_{\cal A}} \, 
\frac{F(\xi_{m+n}, y, x_1,  \ldots , x_m, {\vec x})}
{F(\xi_m, y, x_1,  \ldots , x_m)}
\label{R}
\end{equation}
and $p_{\cal A}$ and $p_{\cal B}$ are the probabilities of
addressing to the sub-processes $\cal A$ and $\cal B$, which, in
principle, may differ.

For solving the polaron problem it is sufficient to have only one pair 
of complementary processes of type II: the sub-process
${\cal A}$ adding a new phonon propagator to the diagram, and its 
counterpart ${\cal B}$ removing one phonon propagator from the 
diagram. 

Consider the algorithm for the process ${\cal A}$. First we select
the position $\tau_1$ for the left-hand end of the extra phonon propagator. 
This is done by choosing at random (with equal probabilities) one of the 
free-electron propagators, and by taking for $\tau_1$ any time (with
equal probability density) within this propagator. Then we select the
position $\tau_2$ for the right-hand end of the phonon propagator,
in accordance with the distribution function 
$\propto \exp [- \omega_p (\tau_2 - \tau_1)]$. After that, we select
the momentum for this propagator, using the distribution
$\propto (1+q/q_0)^{-2}$, where $q_0^2/2=\omega_p$. Now the proposing 
stage is completed, and we are ready to perform accept/reject step,
following the above prescription, Eq.\ (\ref{P_acc}). The corresponding
function $W({\vec x})$ (${\vec x} \equiv \{ \tau_1, \tau_2, {\bf q} \}$)
reads
\begin{equation}
W({\vec x}) \, \propto \, \frac{1}{\tau_0} \,
\frac{1}{(1+q/q_0)^2} \, e^{- \omega_p (\tau_2 - \tau_1)} \; , 
\label{W}
\end{equation}
where $\tau_0$ is the length of the free-electron propagator, where
the point $\tau_1$ is selected. As mentioned earlier, this
form of $W$ is by no means the unique one. Apart from the factor
$p_{\cal B} / p_{\cal A}$ which will be discussed later, the ratio
(\ref{R}) is now completely defined. 

Consider now the algorithm for the process ${\cal B}$. 
We simply select at random (with equal probabilities) some phonon
propagator and with the probabilities given in Eqs.\ (\ref{P_rem}), 
(\ref{W}) remove it.

To complete the description of the sub-processes $\cal A$ and $\cal B$,
we should define the ratio $p_{\cal B} / p_{\cal A}$. It is quite 
reasonable to address to the creation and annihilation procedures
with equal probabilities. At the first glance it might seem that
this immediately leads to $p_{\cal B} / p_{\cal A} = 1$,
but this is not true. The point is that when we select an electron
propagator for placing the point $\tau_1$, we have $N_e$ equal
chances, where $N_e$ is the number of free-electron propagators
in the diagram being modified [denominator of Eq.\ (\ref{R}) ], 
and when we select a phonon propagator for removing, we have $N_{ph}$ 
equal chances, where $N_{ph}$ is the number of phonon propagators 
in the diagram from which we try to remove the propagator 
[numerator of Eq.\ (\ref{R}) ]. These $N_e$ and $N_{ph}$ are 
straightforwardly related to each other:
\begin{equation}
N_{ph} = (N_e +1)/2    \; .
\label{rel}
\end{equation}
We thus get
\begin{equation}
\frac{p_{\cal B} }{ p_{\cal A}}  = \frac{N_e +1}{2N_e} =
\frac{N_{ph}}{2N_{ph} - 1}    \; .
\label{p_ratio}
\end{equation}

As for the processes of type I, these may include (i)
selection of the time $\tau$ anywhere on the interval $(\tau_{2N_{ph}} ,
\infty )$ according to the simple exponential distribution of 
$G^{(0)}({\bf k}, \tau - \tau_{2N_{ph}} )$ [obviously, the role of
the chemical potential is to make this distribution normalizable, and
the whole diagram not to diverge to $\tau \to \infty$; in fact, we
use $\mu$ as a tuning parameter to probe different time-scales since
the typical length of the diagram in time is controlled by the inverse of
$E(k)-\mu $ ], and (ii) the 
change of the diagram momentum from ${\bf k}$ to  ${\bf k+p}$
according to the distribution function 
$\exp [-(\bar{\bf k}+{\bf p})^2 \tau/2m ]$, where $\bar{\bf k}$ is the 
average electron momentum of the diagram, 
i.e., $\bar{\bf k} =\tau^{-1} \int_0^\tau d\tau ' \; 
{\bf k}(\tau ')$. We find it more convenient however, to select the
incoming momentum at will and keep it fixed, since in this case we
collect all the statistics to the value of $k$ we are interested in,
instead of spreading it over the entire $k$-histogram. 
\vspace{1cm}
\begin{figure}
\begin{center}
\epsfxsize=0.15\textwidth
\epsfbox{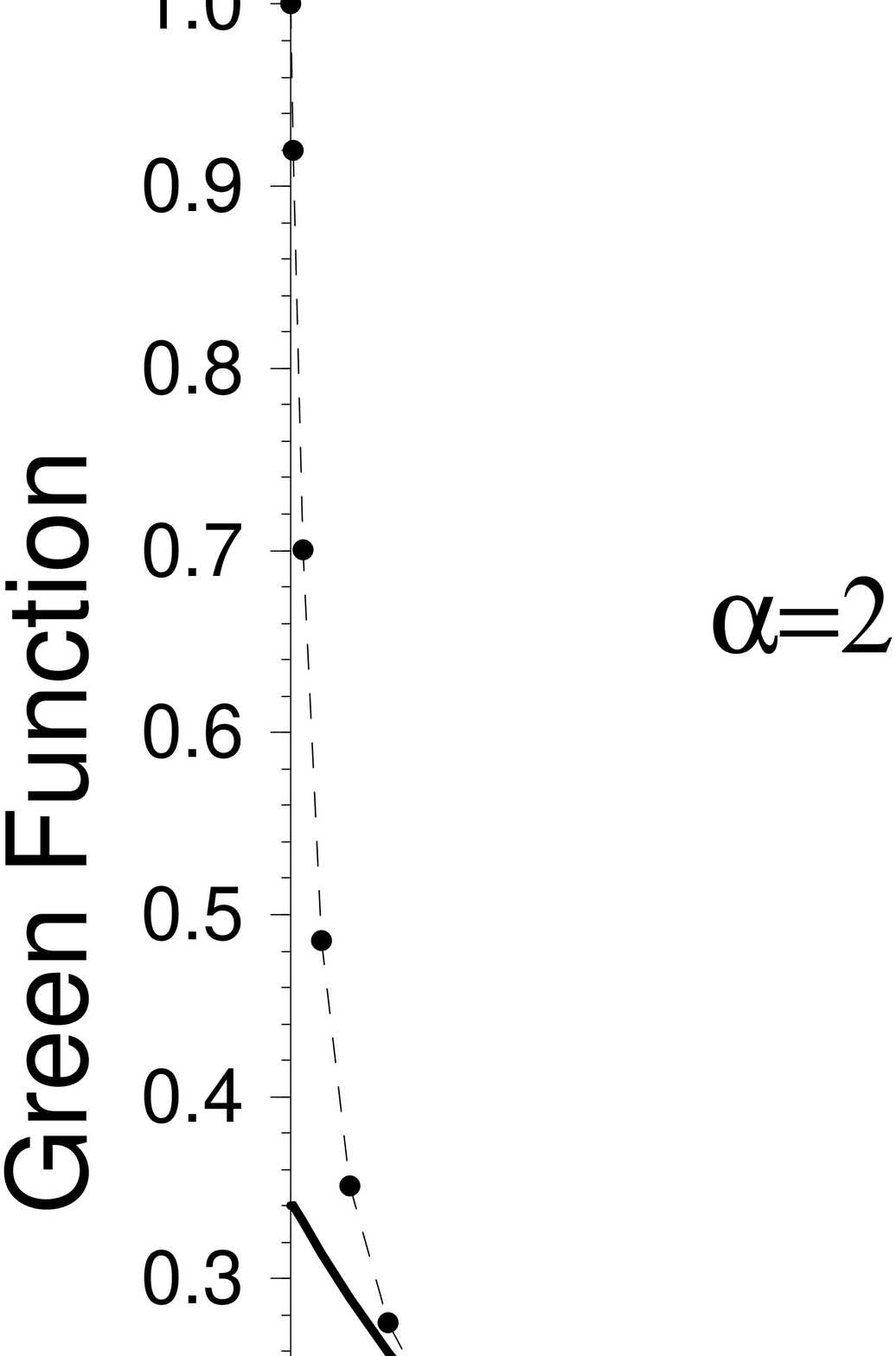}
\epsfxsize=0.15\textwidth
\epsfbox{fig2in.eps}
\end{center}
\caption{Polaron Green function $G(k=0,\tau )$ 
for $\alpha =2$ and $\mu =-2.2$. Solid line is the exponential fit.}
\label{fig:Green}
\end{figure}

In Fig.\ 2 we show the typical data for the polaron Green function.
Following Ref.\cite{Feynman-book}, we use energy units such that
$\omega_p =1$. After initial drop at short times we observe a pure
exponential decay of $G({\bf k}, \tau )$ at longer times 
(provided we are below the threshold of Cherenkov radiation,
$E(k)-E_0 < \omega_p$, so that the polaron state is stable).
From the exponential asymptotic of the Green function we readily
extract the polaron energy:
\begin{equation}
G(k,\tau \gg \omega_p^{-1} ) \, \to \, Z_k \exp [ -(E(k)-\mu ) \tau ] \; .
\label{asym}
\end{equation}
By fine-tuning the chemical potential
very close to $E(k)$ we may extend the time-scale for $G(k,\tau )$ which
is given by $1/(E(k)-\mu )$. Typically, we had reliable statistics
on the time-scale of order $100/\omega_p$, and were thus able to deduce the
polaron energy to accuracy better than $0.01 \omega_p$.
Apart from the polaron energy, the asymptotic behavior of the Green function
(\ref{asym}) gives us one more important physical characteristic of
the polaron, the factor $Z_k$, which shows the fraction of the 
bare-electron state
in the true eigenstate of the polaron:
\begin{equation}
Z_k \, = \, \vert \langle \mbox{free particle~}_k \vert \mbox{~polaron~}_k 
\rangle \vert^2 \; .
\label{Z}
\end{equation}

\begin{figure}
\begin{center}
\epsfxsize=0.27\textwidth
\epsfbox{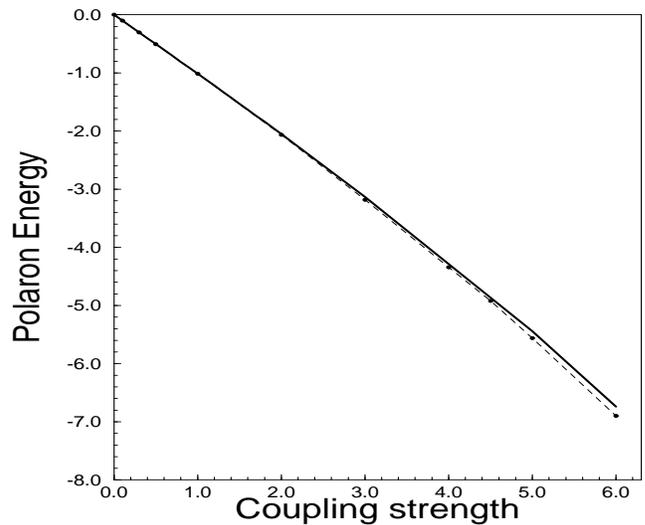}
\end{center}
\caption{Polaron energy $E_0$ as a function of coupling strength. 
Solid line is the Feynman's variational result.}
\label{fig:energy}
\end{figure}

In Fig.\ 3 we present our results for the bottom of the band 
$E_0$ as a function of the coupling strength $\alpha $ in the most
interesting intermediate region $0 < \alpha \le 6 $. As expected,
our precise data are below the solid line which gives the upper bound
for $E_0$ (known to be the lowest ever obtained
for this problem) as derived from the Feynman's variational treatment 
\cite{Feynman-book}. We cannot but note the remarkable
accuracy of the Feynman's approach to the polaron energy.
 
However, the most interesting and instructive data are for the polaron 
spectrum at relatively large $k$. The perturbation theory result for the
dispersion law, 
$E(k) \approx k^2/2 -\alpha (\sqrt{2}/k) \sin^{-1}(k/\sqrt{2})$ (the solid line
in Fig.\ 4),
 clearly demonstrates  that the first-order correction is singular
near the threshold of emitting the optical phonon and even develops
an unphysical maximum (by assumption, the threshold point 
was defined as $E(k_c)=E_0+\omega_p$; the maximum on the dispersion curve
at $k<k_c$ is in contradiction with this assumption). One
is bound to admit then that near the threshold the perturbation theory
for the Fr\"ohlich model fails at any $\alpha $ because of the
singular phonon density of states, which is $\delta $-functional
when one ignores the curvature of the phonon dispersion law
$\omega_p (q) \approx \omega_p =const$ \cite{Whitfield,Larsen}
The formalism dealing
with such cases was developed by Pitaevskii \cite{Pitaevskii} for the
ending point in $^4$He (a similar approach based on the Tamm-Dankoff
approximation was suggested
in Refs.\ \cite{Frohlich,Pines} 
and developed further in \cite{Whitfield,Larsen}). 
By applying it to the Fr\"ohlich model we
arrive at the following equation for the dispersion law
\begin{equation}
\tilde{\omega } -a(k-k_c) + b \tilde{ \omega } 
\int {dx \over x^2-\tilde{\omega }} + R(k-k_c,\tilde{ \omega }) \, = \, 0 \; ,
\label{theory}
\end{equation}
where $\tilde{\omega} \equiv w-(E_0+\omega_p)$, $R$
is a smooth function of $k-k_c$ and $\tilde{\omega }$, and $a$ and $b$ are
some coefficients depending on $\alpha$ and $k_c$. This equation
features an ending point at $k_c$, with the parabolic
dependence Eq.~(\ref{ending-point}) at $k<k_c$. The Monte Carlo data
obtained for $\alpha =1$ are shown in Fig.4. We see how an almost
perfect agreement with the perturbation theory for the band bottom
transforms into non-perturbative behavior near $k_c$ predicted by 
Eq.~(\ref{ending-point}).

\begin{figure}
\begin{center}
\epsfxsize=0.25\textwidth
\epsfbox{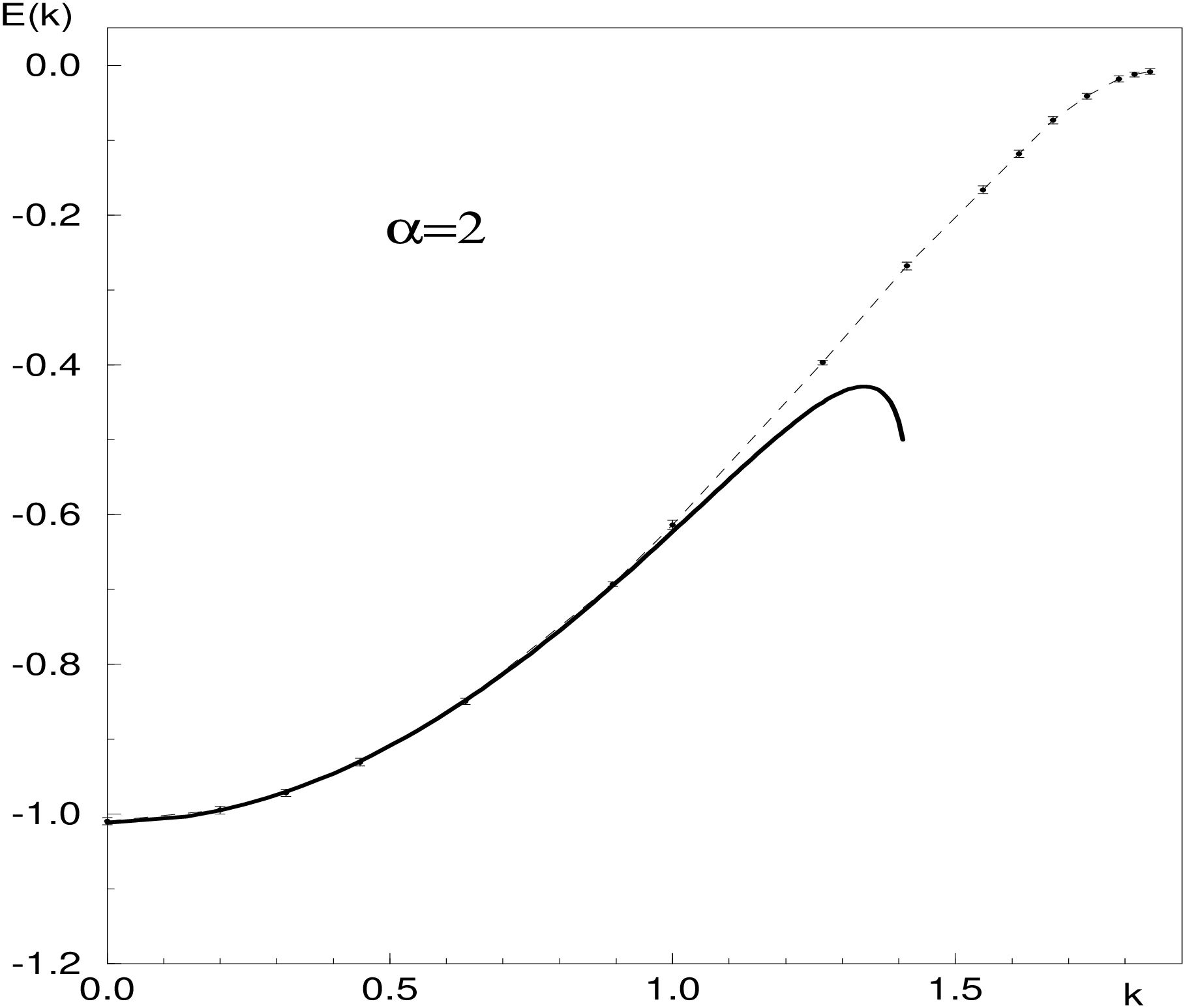}
\end{center}
\caption{Polaron dispersion law for $\alpha =1$. Solid line
is the first-order perturbation theory result.}
\label{fig:spectrum}
\end{figure}

Apparently, the ending-point is an artifact of the dispersionless
phonon spectrum. With the non-zero curvature of $\omega_p(q)$ being
taken into account, the ending point will transform into sharp
crossover from zero to finite damping of the polaron state.
We estimate the crossover region as $\Delta k /k_c \sim \sqrt{m/M}$
where $M$ is the mass of the host-lattice atoms.

In summary, we have presented the solution of the polaron problem by 
the diagrammatic quantum Monte Carlo method. The method
directly simulates the polaron Green function (in the 4D   
momentum-time continuum), dealing with its conventional diagrammatic
expansion. The polaron spectrum, extracted from the asymptotic
behavior of the Green function, demonstrates essentially 
non-perturbative behavior at sufficiently large momenta. 

It is worth noting that diagrammatic Monte Carlo
approach applies to any model dealing with one/few degrees of freedom,
either continuous or discrete, coupled to the thermal bath. Series of
the form Eq.\ (\ref{main}) naturally represent the partition function in the
interaction picture \cite{our96,our98}, and this approach was used
recently to calculate the smearing of the Coulomb staircase in
quantum dots at strong tunneling conductance \cite{german-our}.
More generally, diagrammatic Monte Carlo solves any problem which can
be reduced to Eq.\ (\ref{main}). The efficiency, however, severely depends
on the sign problem, and the convergence becomes very poor if
$F$-functions are not positive definite. It is thus of crucial
importance to work in the representation 
in which the sign problem is absent.

We would like to thank N.\ Nagaosa, A.\ Furusaki, and Yu.\ Kagan for many
fruitful discussions. We also would like to mention that the polaron
problem was brought to our attention and suggested for solution 
by N.\ Nagaosa. We acknowledge the support from the Russian 
Foundation for Basic Research (Grant 98-02-16262).


\begin{references}

\bibitem{Landau33} L.D. Landau, Sov. Phys., {\bf 3}, 664 (1933).

\bibitem{Frohlich} H. Fr$\ddot{\mbox{o}}$hlich, H. Pelzer, and S. Zienau,                 
                     Phil. Mag., {\bf 41}, 221 (1950).

\bibitem{Feynman-book} R.P. Feynman, Statistical Mechanics, Reading,
                       Benjamin (1972); see also R.P. Feynman, Phys. Rev.,
                       {\bf 97}, 660 (1955); T.D. Schultz, Phys. Rev.,
                       {\bf 116}, 526 (1959).

\bibitem{our96}  N.V.\ Prokof'ev, B.V.\ Svistunov, and I.S.\ Tupitsyn,
                 Pis'ma v Zh. Eksp. Teor. Fiz. {\bf 64}, 853 
                 [Sov. Phys. JETP Lett. {\bf 64}, 911] (1996).

\bibitem{our98}  N.V.\ Prokof'ev, B.V.\ Svistunov, and I.S.\ Tupitsyn,
                 to appear in Zh. Eksp. Teor. Fiz. (May 1998);
                 cond-mat/9703200.

\bibitem{Whitfield} G.D. Whitfield and R. Puff, 
     in {\it Polarons and Exitons}, eds. C.G. Kuper and G.D. Whitfield,
     Plenum Press, N.Y., 171 (1962);
     Phys. Rev., {\bf 139}, A338 (1965).

\bibitem{Larsen} D.M. Larsen, Phys. Rev., {\bf 144}, 697 (1966).

\bibitem{Pitaevskii} L.P. Pitaevskii, Zh. Eksp. Teor. Fiz., {\bf 36},
                     1168 (1959). 


\bibitem{Pines} D. Pines, 
     in {\it Polarons and Exitons}, eds. C.G. Kuper and G.D. Whitfield,
     Plenum Press, N.Y., 155 (1962).

\bibitem{german-our} G. G\"{o}ppert, H. Grabert, N. Prokof'ev, and
      B.V. Svistunov, submitted to Phys. Rev. Lett. (February 1998),
      cond-mat/9802248.

\end{references}
\end{document}